\documentstyle[multicol,aps,prl,graphicx,amssymb]{revtex}
\draft

\begin{document}

\title{Bridging the ARCH model for finance and nonextensive entropy}
\author{S\'{\i}lvio M. Duarte Queir\'{o}s and Constantino Tsallis\thanks{%
sdqueiro@cbpf.br, tsallis@cbpf.br}}
\address{Centro Brasileiro de Pesquisas F\'{\i}sicas, Rua Dr. Xavier Sigaud 150,\\
22290-180, Rio de Janeiro-RJ, Brazil}
\date{\today}
\maketitle

\begin{abstract}
Engle's ARCH algorithm is a generator of stochastic time series for
financial returns (and similar quantities) characterized by a time-dependent
variance. It involves a memory parameter $b$ ($b=0$ corresponds to {\it no
memory}), and the noise is currently chosen to be Gaussian. We assume here a
generalized noise, namely $q_n$-Gaussian, characterized by an index $q_{n}
\in {\cal R}$ ($q_{n}=1$ recovers the Gaussian case, and $q_n>1$ corresponds
to tailed distributions). We then match the second and fourth momenta of the
ARCH return distribution with those associated with the $q$-Gaussian
distribution obtained through optimization of the entropy $S_{q}=\frac{%
1-\sum _{i} {p_i}^q}{q-1}$, basis of nonextensive statistical mechanics. The outcome is an {\it analytic} distribution for the returns, where
an unique $q\ge q_n$
corresponds to each pair $(b,q_n)$ ($q=q_n$ if $
b=0$). This distribution is compared with numerical results and appears to
be remarkably precise. This system constitutes a simple,
low-dimensional, dynamical mechanism which accommodates well within the
current nonextensive framework.
\end{abstract}

\pacs{PACS numbers: 05.40.-a, 05.90.+m, 89.65.Gh}

\begin{multicols}{2}

Time series are ubiquitous in nature. They appear in geoseismic phenomena, El Ni\~no, finance, electrocardio- and electroencephalographic profiles, among many others. Some of these series can be constructed by using, for successive values ot time $t$, random variables associated with the {\it same} distribution for all times: they are called {\it homoskedastic}. Such is the case of the ordinary Brownian motion. But many phenomena exist in nature which do {\it not} accomodate with such a property, i.e., the distribution associated with each value of $t$ depends on $t$. Such random variables are then called {\it heteroskedastic}. A simple illustration would be to use say a centered Gaussian at all steps, but with a {\it randomly varying width}. 

Nowadays, time series that are intensively studied are the financial ones, where
nonstationary {\it volatility} (technical name for second-order
moment of say returns) is a common feature \cite{mandelbrot,fama,bouchaud}. In order to mimic explicitly and analyze this type of time series, R.F. 
Engle introduced in 1982 the {\it autoregressive conditional heteroskedasticity} (ARCH) process\cite{engle}. Its prominence can be measured by its wide use, by the amount of its extensions introduced later \cite{boller,granger}, and --- last but not least --- by the fact that the 2003 Nobel Prize for Economics was awarded to Engle ``for methods of analyzing economic time series with time-varying volatility (ARCH)". This and similar processes are commonly used in the implementation of asset pricing theories, market
microstructure models, and pricing derivative assets \cite{granger,mccurdy} (see also \cite{farmer}). Specifically, the ARCH($s$)\cite{engle} process
generates, for the {\it returns} $z_{t}$, a discrete time series  whose variance 
$\sigma _{t}^{2}$, at each time step, depends {\it linearly} on the $s$ {\it previous}
values of $[z_{t}]^2$. It is defined as follows:

\begin{equation}
\left\{ 
\begin{array}{c}
z_{t}=\sigma _{t} \;\omega _{t}, \\ 
\sigma _{t}^{2}=a+   \sum_{i=1}^s    b_i \; z_{t-i}^{2} \;,
\end{array}
\right.   \label{def-arch}
\end{equation}
or, equivalently,
\begin{equation}
\left\{ 
\begin{array}{c}
z_{t}=\sqrt{a+  \sum_{i=1}^s b_i \; z_{t-i}^{2}}\,\;\omega _{t}  \\
\sigma _{t}^{2}=a+ \sum_{i=1}^s b_i \;   \sigma^2_{t-i}\;\omega^2_{t-i} \; ,  
\end{array}
\right.   \label{def-arch2}                       
\end{equation}
where $a,b_{i},\sigma _{t}^{{}}\geq 0$, and $\omega _{t}$ 
represents an independent and identically distributed
stochastic process with mean value null and unit variance, (i.e., $\left\langle
\omega _{t}\right\rangle =0$ and $\left\langle \omega _{t}^{2}\right\rangle
=1$), currently chosen to follow a Gaussian distribution, but other choices
are possible. In this manuscript we will discuss in detail the usual case $s=1$,
ARCH(1), which will be from now on simply designated by ARCH. In general, the
distribution, $P_{n}\left( \omega \right) $, together with parameters $a$
and $\{b_i\}$, specify the particular ARCH\ process ($n$ stands for {\it noise}).

As can be seen from Eqs. (\ref{def-arch2}), parameters $\{b_i\}$ characterize the memory
mechanism. For $b_i=0 \; (\forall i)$,
there is no memory effect, and consequently the ARCH process for returns 
reduces to generating the noise $\omega _{t}$ (multiplied by $\sqrt{a}$). We can verify that, in general, $\langle z_t \rangle = \langle \sigma_t \rangle \langle \omega_t \rangle=0$, and $\langle z_t \,z_{t^\prime} \rangle = \langle \sigma_t \, \sigma_{t^\prime} \rangle \langle \omega_t \, \omega_{t^\prime} \rangle=\langle \sigma_t \, \sigma_{t^\prime} \rangle \,\delta_{tt^\prime}$. Therefore, even for nonvanishing $b_i$'s, the returns
$z_{t}$ remain {\it uncorrelated}, whereas correlations do exist in the variance $\sigma _t^2$ \cite{boller}. We observe that this 
stochastic process captures the
tendency for the so called {\it volatility clustering}, i.e.,  large (small) values
of $z_{t}$ are followed by other large (small) values of $z_{t+1}$, {\it but of
arbitrary sign}. In other words, $\langle |z_t| \, |z_{t^\prime}| \rangle$ by no means is proportional to 
$\delta_{tt^\prime}$. 

Hereon we focus on ARCH(1), i.e., $b_1\equiv b$ and $b_i=0\;(i=2,\,3,...)$. 
From Eqs. (\ref{def-arch}) it is simple to
obtain the n-th moment for the $P(z)$ stationary distributions, particularly the second
moment
\begin{equation}
\sigma^{2} \equiv \left\langle z_{t}^{2}\right\rangle  = \langle \sigma_t^2 \rangle =\frac{a}{1-b}\;\;\;
(b<1),  \label{z2}
\end{equation}
and the fourth moment,
\begin{equation}
\left\langle z^{4}\right\rangle =a^{2}\left\langle \omega_t^{4}\right\rangle 
\frac{1+b}{\left( 1-b\right) \left( 1-b^{2}\left\langle \omega_t^{4}\right\rangle \right) }\,.
\end{equation}

Without loss of generality, we can assume that the ARCH procedure
generates a time series with unit variance, i.e., $\sigma^{2}=1$, hence 
$a=1-b$. Now, for $z_t$, the fourth moment is numerically equal to the {\it kurtosis} $k_{x} \equiv \frac{\left\langle x^{4}\right\rangle }{\left\langle
x^{2}\right\rangle ^{2}}$ (a
possible measure of non-gaussianity or peakedness for probability
distributions), and we can easily get,
 
\begin{equation}
\left\langle z_t^{4}\right\rangle =k_{z}=k_{\omega }\Bigl[1+b^{2}\frac{%
k_{\omega }-1}{1-k_{\omega }b^{2}}\Bigr]  \;\;\;\;( k_{\omega }b^{2}<1)\;, \label{k}
\end{equation}
It is straightforwardly verified that, whatever the
form of $P_{n}\left( \omega \right) $, the ARCH process generates
probability distributions $P\left( z\right) $ with a {\it slower} decay and
consequently with a kurtosis $k_{z}>k_{w}$ \cite{white,weiss,boller2,stanley}. See in Figs. 1 and 2, typical runs for a Gaussian noise. 

\begin{figure}[tbp]
\begin{center}
\includegraphics[width=7cm,angle=0]{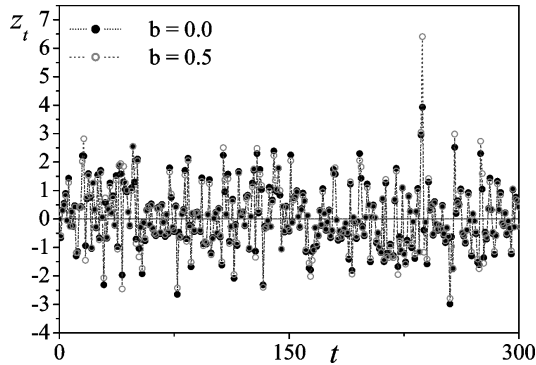}
\end{center}
\caption{{{{\protect\small Two examples of ARCH time series obtained for a {\it Gaussian} noise (i.e., $q_n=1$).
Memory (i.e., $b>0$) increases the probability for large values of $|z_{t}|$, i.e., fat tails in $P(z)$. The large value at $t=237$
is  virtually never observed for $b=0$. 
}}}}
\label{z-vs-t}
\end{figure}

\begin{figure}[tbp]
\begin{center}
\includegraphics[width=7cm,angle=0]{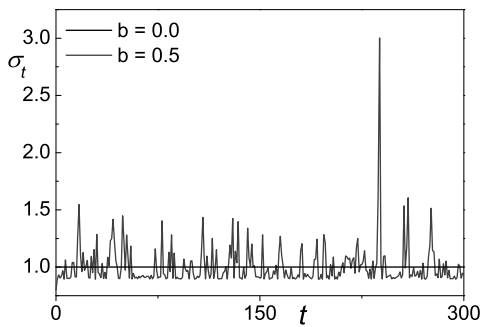}
\end{center}
\caption{{{{\protect\small Time dependence of the volatility $\sigma$ for the same two ARCH processes shown in 
Fig. \ref{z-vs-t}. The largest value for $\sigma$ occurs at $t=238$. }}}}
\label{vol-vs-t}
\end{figure}

We shall now establish a possible
analytical connection between the memory parameter $b$, $P\left( z\right) 
$ and $P_{n}\left( \omega \right) $ under the framework of the nonextensive statistical mechanics introduced by one of us \cite{tsallis}.
Consider a probability distribution $P^{\prime }\left( z\right) $ such that
it maximizes the entropic form 
\begin{equation}
S_{q}=\frac{1-\int_{-\infty }^{+\infty }\left[ P^{\prime }(z)\right] ^{q}dz}{q-1}\;\;\;\;(q \in \ \Re )\,,
\label{sq}
\end{equation}
with $\lim_{q\to 1} S_q=-\int_{-\infty }^{+\infty }P^{\prime }(z)\ln P^{\prime }(z)dz \equiv S_{BG}$ ($BG$ stands for {\it 
Boltzmann-Gibbs}).
Maximizing Eq. (\ref{sq})
with the constraints $\int_{-\infty }^{+\infty }P^{\prime }(z)dz=1$ and$\
\int_{-\infty }^{+\infty }z^{2}P^{\prime }(z)dz=1$ we have

\begin{equation}
P^{\prime }(z)=\frac{{\cal A}}{[ 1+{\cal B}( q-1) z^2] ^{\frac{1}{q-1}}} \;\;\;\;(q<5/3)\,,  \label{pz}
\end{equation}
with
\begin{equation}
{\cal B} \equiv \frac{\Gamma \left[ \frac{5-3q}{2q-2}\right] }{%
2(q-1)\Gamma \left[ \frac{3-q}{2q-2}\right] }\,,  \label{calb}
\end{equation}
and
\begin{equation}
{\cal A} \equiv \frac{\Gamma \left[ \frac{%
1}{q-1}\right] }{\sqrt{2\pi }\Gamma \left[ \frac{3-q}{2q-2}\right] }\sqrt{(2q-1){\cal B}}\,.  \label{cala}
\end{equation} 
In the $q\rightarrow 1$ limit,  the normal distribution is recovered. The fourth moment
(which for this case coincides with the kurtosis) is given by, 
\begin{equation}
\left\langle z^{4}\right\rangle =3\frac{\Gamma \left[ \frac{7-5q}{2q-2}%
\right] \Gamma \left[ \frac{3-q}{2q-2}\right] }{\left\{ \Gamma \left[ \frac{%
5-3q}{2q-2}\right] \right\} ^{2}} \;\;\;\;( 1<q<\frac{7}{5} )  \;. \label{kq}
\end{equation}
Let us now make the {\it ansatz} $P^{\prime }\left(
z\right)  \simeq $ $P\left( z\right) $. Consistently we impose the matching of Eqs. (\ref{k})
and (\ref{kq}) . This yields $q$ as a function of $b$ and $k_\omega$.
Assuming that noise $\omega _{t}$ follows a
generalized distribution, Eq. (\ref{pz}), defined by an entropic index $q_{n}$ (such that $k_{w}^{(q_{n})}b^{2}<1$) we are able to establish a relation
between the memory parameter $b$ and the entropic indexes $q$ and $q_{n}$, which characterize respectively the distributions $P^\prime\left( z\right) $ and $%
P_{n}\left( w\right) $. This connection is straightforwardly given by
\begin{equation}
b=\frac{\sqrt{G(q)\left\{ \Gamma \left[ \frac{5-3q_{n}}{2q_{n}-2}\right]
\right\} ^{2}-G(q_{n})\left\{ \Gamma \left[ \frac{5-3q}{2q-2}\right]
\right\} ^{2}}}{\sqrt{G(q_{n})\left( 3G(q)-\left\{ \Gamma \left[ \frac{5-3q}{%
2q-2}\right] \right\} ^{2}\right) }} \, ,  \label{b}
\end{equation}
where $G(x)\equiv\Gamma \left[ \frac{7-5x}{2x-2}\right] \Gamma \left[ \frac{3-x}{%
2x-2}\right] $. This connection, depicted in Fig. 3, constitutes, to the best of our knowledge, the first ever found which {\it analytically} expresses the return distribution in terms of the noise distribution and the memory parameter $b$. In what follows we shall verify that the above ansatz is indeed satisfied within a remarkable precision. To do this for typical values of $b$ and $q_{n}$, we first generate, through a standard algorithm based on Eq.(\ref{def-arch}) (with $s=1$), a set of ARCH time series and their corresponding {\it probability density functions} (PDF's) $P(z)$. The results are indicated in Figs. 4 and 5. Then we compare with a histogram (with a conveniently chosen unit interval $\delta$) associated with the distribution (\ref{pz}), with $q$ satisfying Eq. (\ref{b}). In other words, we compare with ${\cal A}%
_{\delta }\left[ 1-{\cal B}\left( 1-q\right) x^{2}\right] ^{\frac{1}{1-q}}$,
where ${\cal B}$ is given by Eq. (8), and
\begin{equation}
\begin{array}{c}
{\mathcal{A}}_{\delta }=\int_{-\delta /2}^{+\delta /2}p(z)dz=\frac{\delta }{%
\sqrt{2\pi }}\frac{\Gamma \left[ \frac{1}{q-1}\right] }{\Gamma \left[ \frac{%
3-q}{2q-2}\right] }\sqrt{\frac{\Gamma \left[ \frac{5-3q}{2q-2}\right] }{%
\Gamma \left[ \frac{3-q}{2q-2}\right] }}  \\ 
\times \, _{2}F_{1}\left( \frac{1}{2},\frac{1}{q-1};\frac{3}{2},-\frac{\delta
^{2}\Gamma \left[ \frac{5-3q}{2q-2}\right] }{8\Gamma \left[ \frac{3-q}{2q-2}%
\right] }\right) 
\end{array}
\label{ad}
\end{equation}

\begin{figure}[tbp]
\begin{center}
\includegraphics[width=7cm,angle=0]{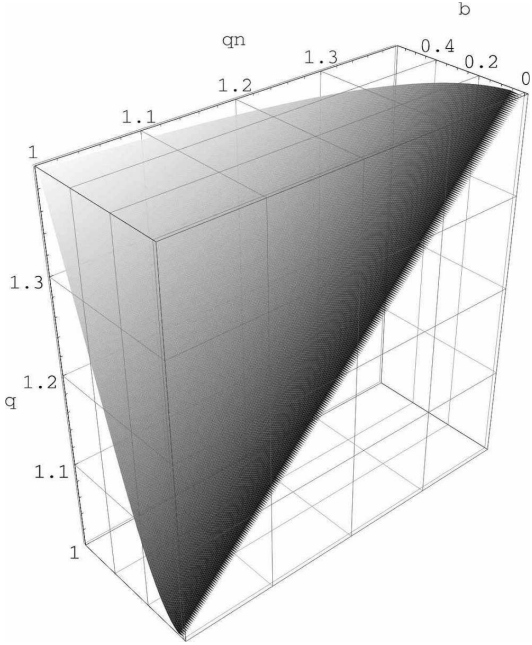}
\end{center}
\caption{ Diagram $(q,q_{n},b)$ for the ARCH process with $\left\langle
z^{2}\right\rangle =1$ (for $b=0$ we have the straight line $q=q_{n}$).}
\label{q-qn-b-gray1}
\end{figure}

\begin{figure}[tbp]
\begin{center}
\includegraphics[width=8.5cm,angle=0]{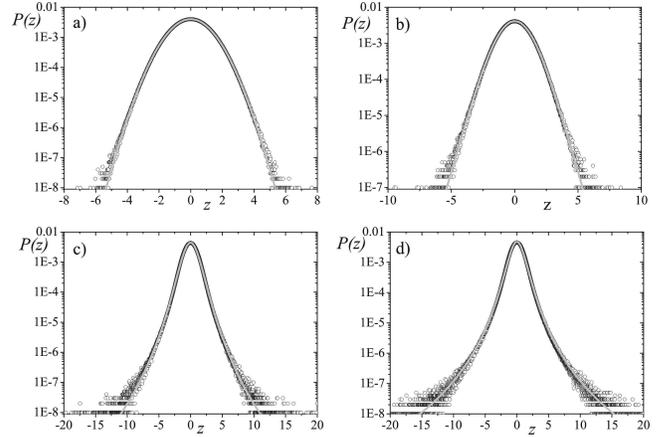}
\end{center}
\caption{{{{\protect\small PDFs for a $q_{n}=1$ noise and typical values 
of $b$. a) $b=0$, $q=1$ ($\protect\chi ^{2}=1.14\times 10^{-11}$); b) $b=0.1$%
, $q=1.01976$ ($\protect\chi ^{2}=2.34\times 10^{-10}$); c) $b=0.4$, $%
q=1.242424$ ($\protect\chi ^{2}=1.82\times 10^{-10}$); d) $b=0.5$, $q=\frac{4}{%
3}$ ($\protect\chi ^{2}=2.98\times 10^{-10}$). }}}}
\label{fig-pdf1}
\end{figure}

\begin{figure}[tbp]
\begin{center}
\includegraphics[width=8.5cm,angle=0]{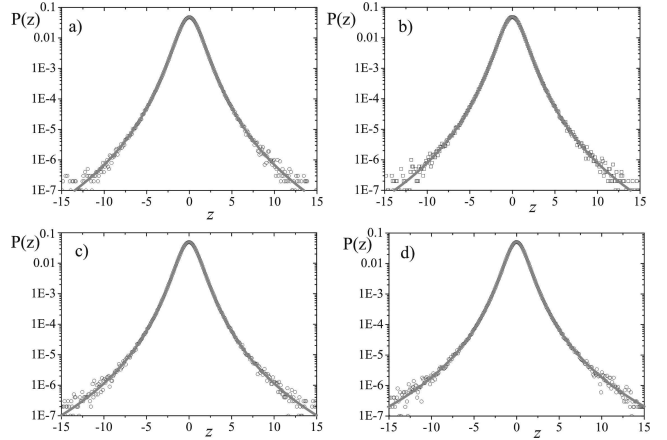}
\end{center}
\caption{{{{\protect\small PDFs for a $q_{n}=1.3$ noise and typical 
values of $b$. a) $b=0$, $q=1.3$ ($\protect\chi ^{2}=3.93\times 10^{-10}$);
b) $b=0.1$, $q=1.30762$ ($\protect\chi ^{2}=8.86\times 10^{-10}$); c) $b=0.2$%
, $q=1.32950$ ($\protect\chi ^{2}=9.72\times 10^{-10}$); d) $b=0.3$, $%
q=1.36306$ ($\protect\chi ^{2}=2.68\times 10^{-10}$). }}}}
\label{fig-pdf13}
\end{figure}

We verify that the agreement is quite satisfactory. In order to quantify the small discrepancy between the ARCH PDF's $P(z)$ and the analytical
ones $P^\prime(z)$, we have indicated in the captions of Figs. \ref{fig-pdf1} and \ref{fig-pdf13} the values 
of $\chi ^{2} \equiv \frac{1}{N} \,  \sum\limits_{i=1}^{N}[P(z_{i})-P^{\prime }(z_{i})] ^{2}$, $N$ being the number of points. 
An alternative evaluation of the discrepancy of the ARCH and analytical distribution can be done by comparing the sixth-order moment. 
It is simple to verify that these momenta for $P(z)$ and $P^{\prime}(z)$ exhibit quite small 
discrepancies. For example, for the cases illustrated in Figs. 
\ref{fig-pdf1} and \ref{fig-pdf13},  we have obtained discrepancies never larger 
than $2.9 \%$ (occuring in fact for $b \simeq 0.5$) for $q_n=1$ and 
than $2.7 \%$ (occuring in fact for $b \simeq 0.3$) for $q_n=1.3$. Such minor 
discrepancies are, in financial practice, completely inocuous (for example, we may 
check in \cite{osorioborlandtsallis,lisa}  that realistic return distributions 
are larger than $10^{-6}$, whereas in our present Figs. \ref{fig-pdf1}  and \ref{fig-pdf13} we have simulated down to $10^{-8}$). 

Let us conclude by saying that the fact that a close connection has been shown to exist between the 
possibly ubiquitous ARCH stochastic processes and the nonextensive entropy  Eq.(\ref{sq}) strongly suggests further possible
connections. We refer to phenomena which might present some kind of scale
invariant geometry, e.g., hierarchical or multifractal structures, 
low-dimensional dissipative and conservative maps\ \cite{mapas}, fractional
and/or nonlinear Fokker-Planck equations \cite{fokker}, Langevin dynamics
with fluctuating temperature \cite{langevin}, possibly scale-free network growth \cite
{network}, long-range many-body classical Hamiltonians \cite{rotores,reviews,GM-CT}, among others.  A more detailed 
study of this as well as of other heteroskedastic models (e.g., GARCH) is in progress.

We acknowledge useful remarks from J.D. Farmer, as well as partial support from Faperj, CNPq, 
PRONEX/MCT (Brazilian agencies) and FCT/MCES (contract SFRH/BD/6127/2001) (Portuguese agency).

\end{multicols}

\end{document}